\documentclass[aps,prl,,showpacs,twocolumn,groupedaddress]{revtex4}
\usepackage{graphicx}
\usepackage{epstopdf}
\begin{document}

\title{Persistent local order heterogeneity in the supercritical carbon dioxide}
\author{Dima Bolmatov$^{1}$}
\thanks{d.bolmatov@gmail.com, db663@cornell.edu}
\author{D. Zav'yalov$^{2}$}
\author{M. Gao$^{3}$}
\author{M. Zhernenkov$^{4}$}
\address{$^1$ Baker Laboratory, Cornell University, Ithaca, New York 14853, USA}
\address{$^2$ Volgograd State Technical University, Volgograd, 400005 Russia}
\address{$^3$ Queen Mary University of London, Mile End Road, London, E1 4NS, UK}
\address{$^4$ Brookhaven National Laboratory, Upton, NY 11973, USA}

\begin{abstract}
The supercritical state is currently viewed as uniform and homogeneous on the pressure-temperature phase diagram in terms of physical properties. Here, we study structural properties of the supercritical carbon dioxide, and discover the existence of persistent medium-range order correlations which make supercritical carbon dioxide non-uniform and heterogeneous on an intermediate length scale, a result not hitherto anticipated. We report on the carbon dioxide heterogeneity shell structure where, in the first shell, both carbon and oxygen atoms experience gas-like type interactions with short range order correlations, while within the second shell oxygen atoms essentially exhibit liquid-like type of interactions with medium range order correlations due to localisation of transverse-like phonon packets. We show that the local order heterogeneity remains in the three phase-like equilibrium within very wide temperature range. Importantly, we highlight a catalytic role of atoms inside the nearest neighbor heterogeneity shell in providing a mechanism for diffusion in the supercritical carbon dioxide on an intermediate length scale. Finally, we discuss important implications for answering the intriguing question whether Venus may have had carbon dioxide oceans and
urge for an experimental detection of this persistent local order heterogeneity.
\end{abstract}

\maketitle
Unlike the long-range order of ideal crystalline structures, local order is an intrinsic characteristic of soft matter materials and often serves as the key to the tuning of their properties and their possible applications \cite{ziman1,mon1,cip1,harr1,chand1,lar1,harr2,col1}. Liquid materials often exhibit medium and/or short range order, such as nanoscale structural motifs, clustering, or local ordering at the level of the atomic coordination spheres. Describing this local order of soft matter systems and supercritical fluids in particular is crucial for understanding the origins of their material properties \cite{pine1,tanaka1,klein1,uhle1,gutt1,howew1}.

Theoretical understanding of the supercritical state is lacking, and is seen to limit further industrial deployment. Supercritical fluids have started to be employed in several important applications, ranging from the extraction of floral fragrance from flowers to applications in food science such as creating decaffeinated coffee, pharmaceuticals, polymers, cosmetics, functional food ingredients, powders, bio- and functional materials, natural products, nano-systems, biotechnology, fossil and bio-fuels, microelectronics, energy and environment. The development of new experimental methods and improvement of existing ones continues to play an important role in this field \cite{kolesov1,huang1,cunsolo1,saito1,bolmatov1,bolmatov2}, with  the ongoing effort in elucidating and understanding the structure and properties of disordered matter such as glasses and liquids \cite{hopkins1,dyre1,hangjun1,jani1,anisimov1,tan2}.

\begin{figure}
\includegraphics[scale=0.34]{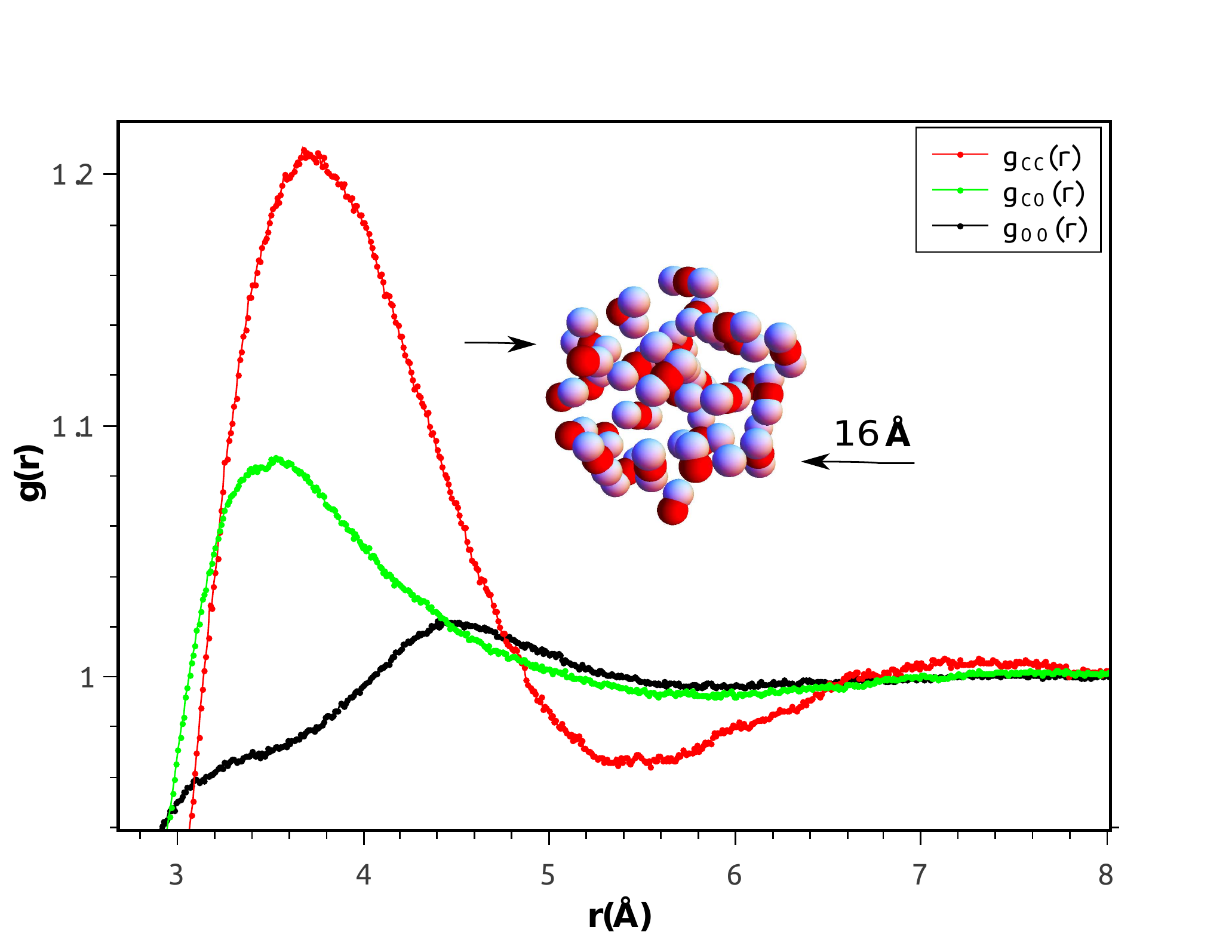}
\caption{ Intermolecular partial pair distribution functions of supercritical CO$_2$. Partial pair distribution functions  and a heterogeneity generated by molecular dynamics simulations at T$=10000$ K. The inserted heterogeneity, consist of carbon atoms (Red) and oxygen atoms (Silver Blue), was depicted from a simulated configuration.}
	\label{fig1}
\end{figure}

Supercritical carbon dioxide has been extensively studied in neutron, x-ray wide and small angle diffraction experiments \cite{rio1,lyn1} but in limited ranges of temperature and pressure therefore their unusual properties at elevated temperatures and pressures have not been amenable to be unveiled. Here, we focus on the structural properties of the supercritical carbon dioxide within very wide temperature range far beyond the critical point. On the basis of molecular dynamics simulations, we calculate partial pair distribution functions and partial static structure factors. A further most fascinating observation is robust medium-range order correlations which make the supercritical CO$_2$ non-uniform on an intermediate length scale. Therefore, the persistent structural local order heterogeneity (see Fig. 1) challenges the currently held belief that the supecritical state is uniform and homogeneous in terms of physical properties \cite{kiran1}.

\begin{figure}[htp]
  \centering
  \begin{tabular}{cc}
    
    \includegraphics[width=43mm]{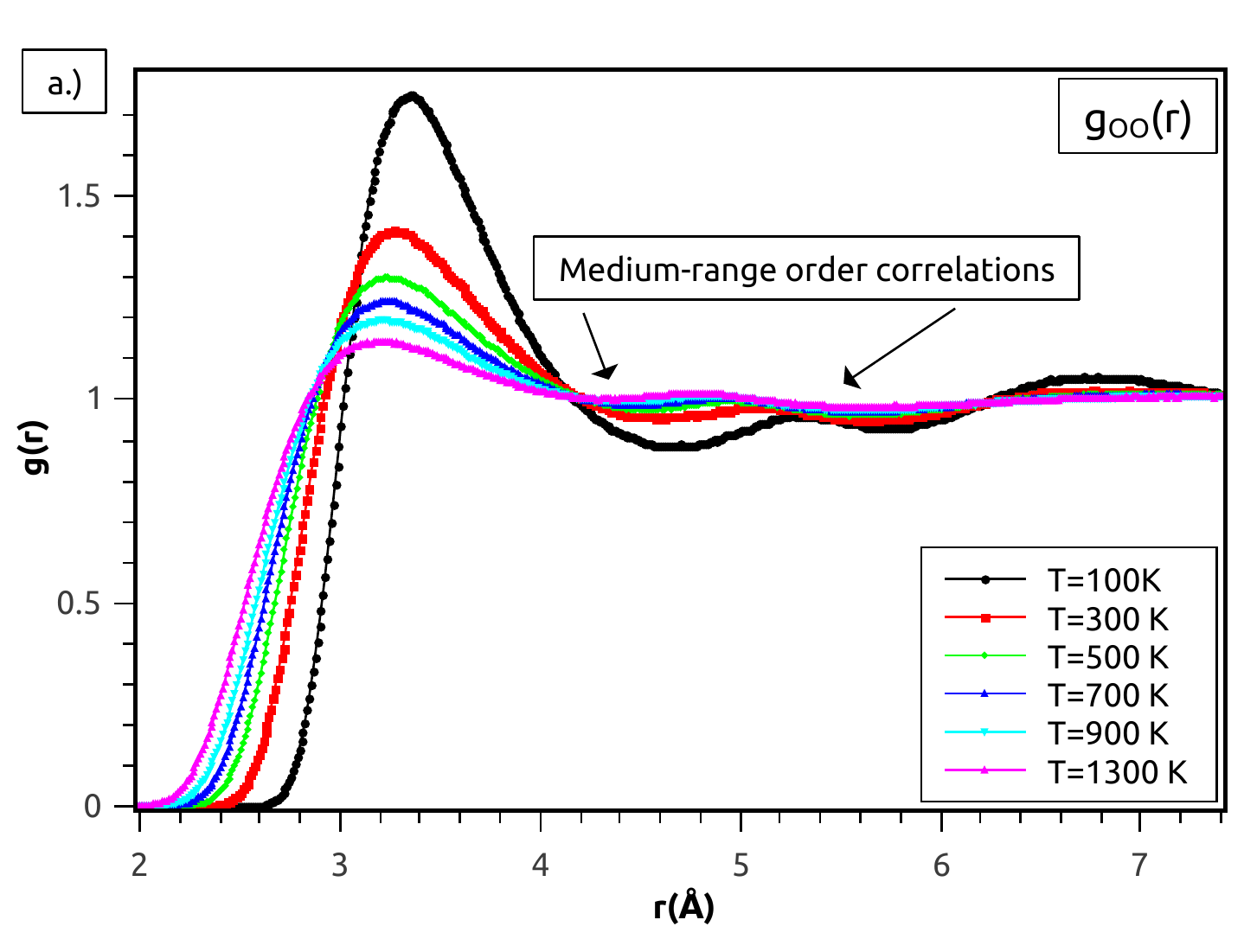}&

    \includegraphics[width=43mm]{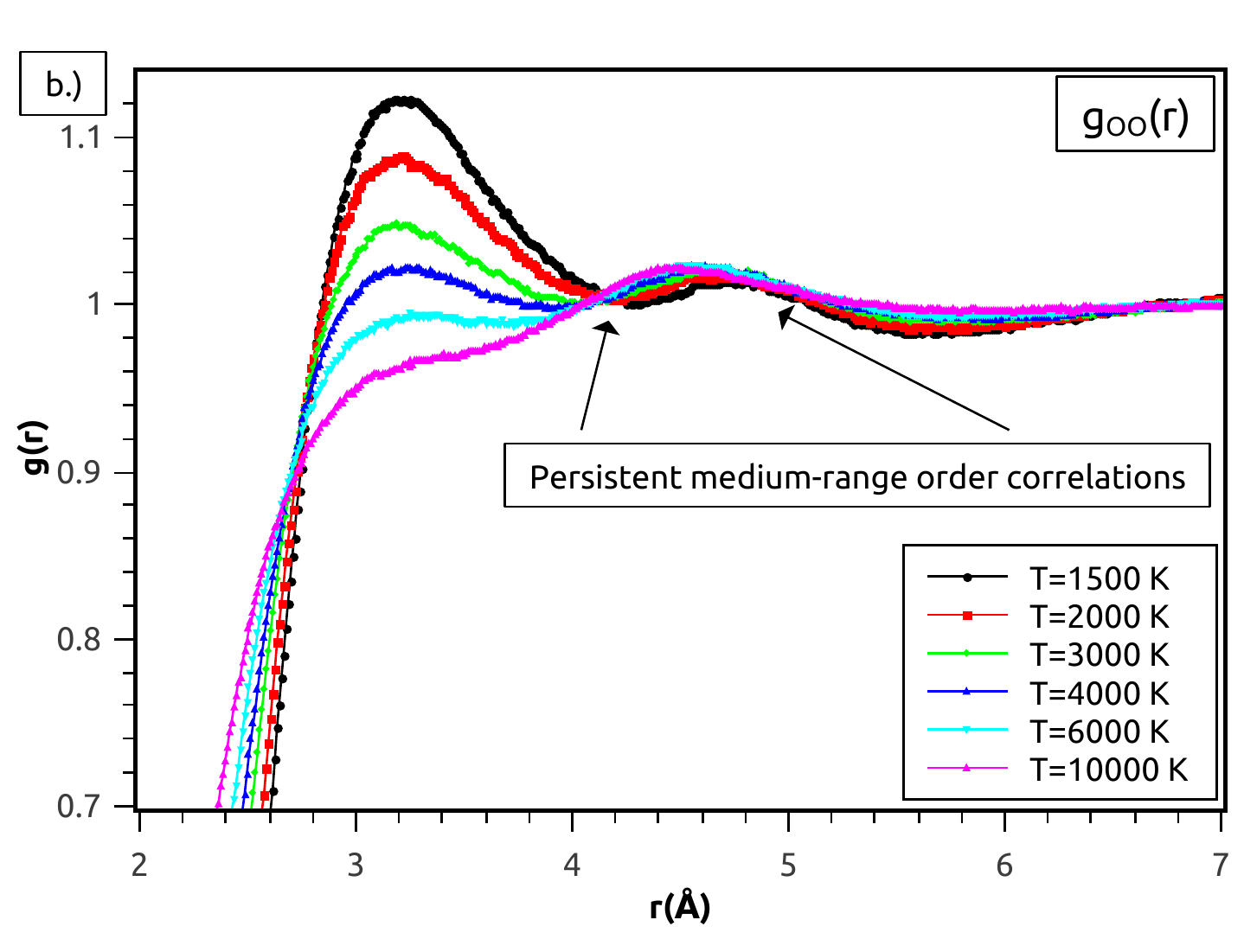}\\

    \includegraphics[width=43mm]{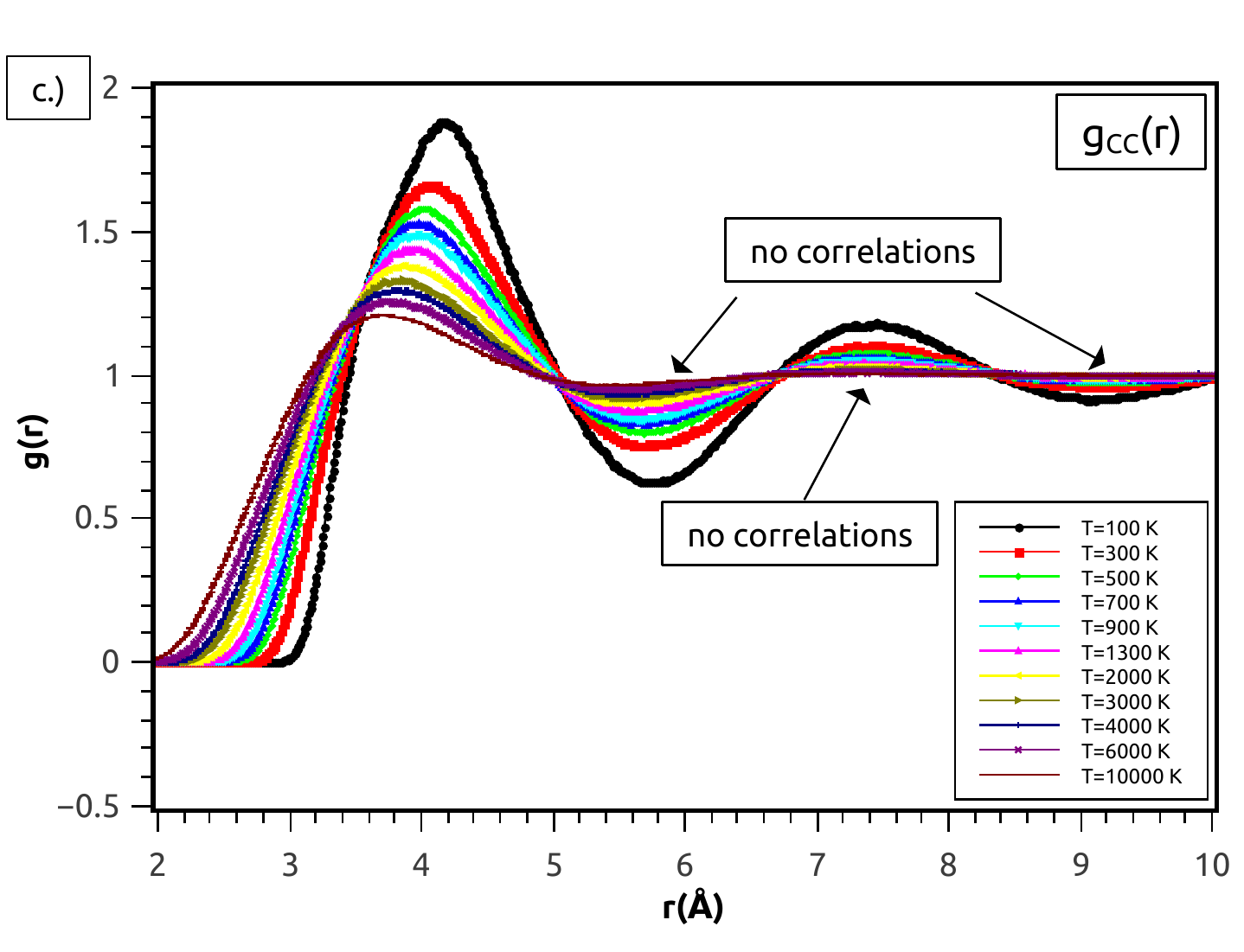}&

    \includegraphics[width=43mm]{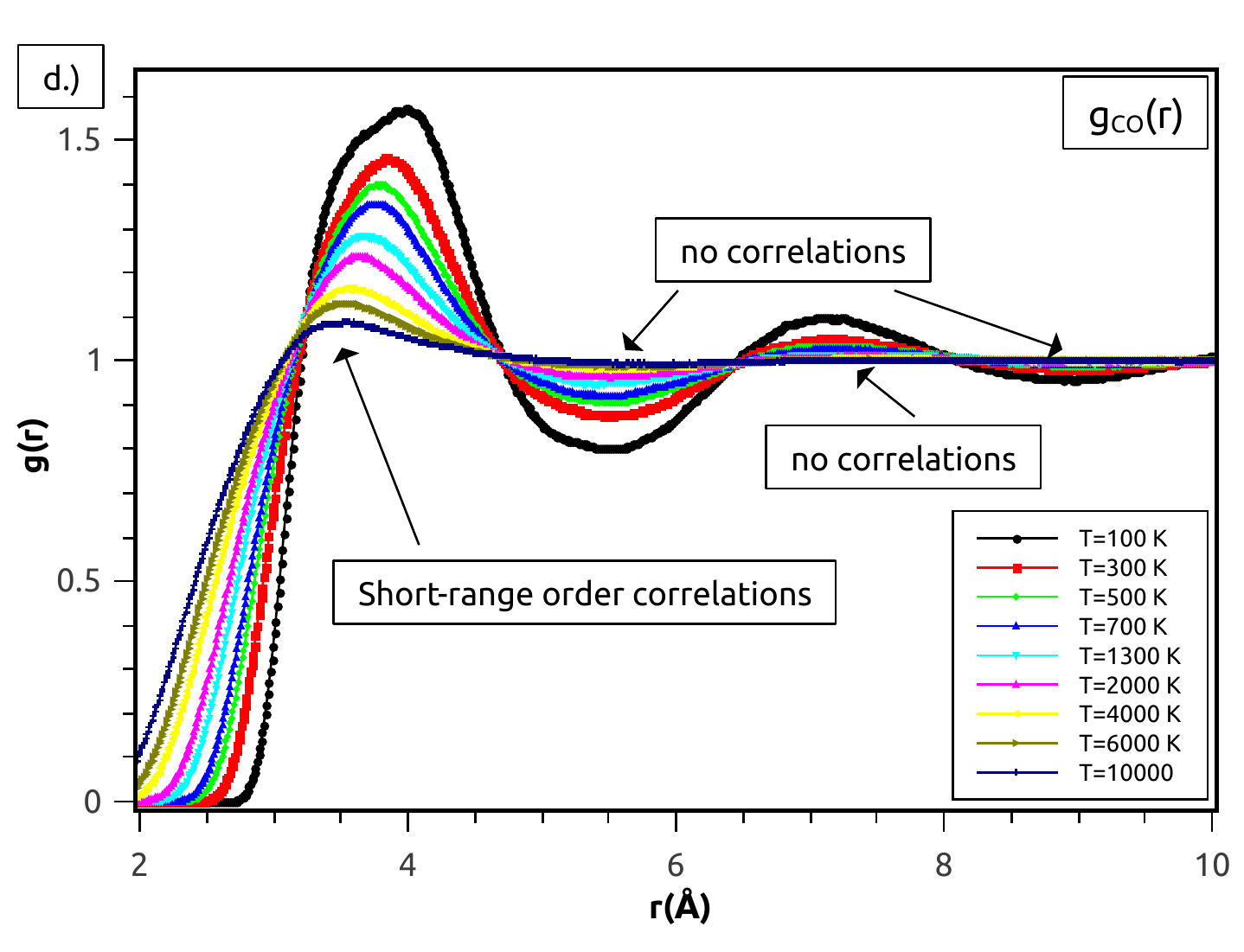}\\

  \end{tabular}
 \caption{ Structural evolution in real space. Evolution of intermolecular partial pair distribution functions. A sharpening of the O--O correlations on an intermediate length scale arising from increased orientational ordering of the supercritical carbon dioxide fluid.}
 \label{figur}
\end{figure}





Using Molecular Dynamics (MD) simulations, we have simulated supercritical carbon dioxide fluid. We have simulated the system with 14000 atoms using constant-volume (NVE) ensemble in the very wide temperature range (see Figs. 2--3) well extending into the supercritical region; the system was equilibrated at constant temperature. The temperature range in Figs. 2--3 is between about T$_c$ and 33T$_c$, where T$_{c}\simeq$ 304.1 K is the critical temperature of CO$_2$. The simulated density, 1195.1 kg/m 3, corresponds to approximately three times the critical density of CO$_{2}$. A typical MD simulation lasted about 40 ps, and the properties were averaged over the last 20 ps of simulation, preceded by 20 ps of equilibration. The simulations at different temperature included 100 temperature points simulated on the high-throughput computing cluster. We calculate pair distribution functions $g(r)$ (PDF) and partial static
structure factors $S(q)$, average it over the last 20 ps of the simulation, and show its temperature evolution. 
Particles in a gas move in almost straight lines until they collide with other particles or container walls and change course. In liquids, particle motion has two components: a solid-like, quasi-harmonic vibrational motion about equilibrium locations and diffusive jumps between neighboring equilibrium position. As the temperature increases or the pressure decreases, a particle spends less time vibrating and more time diffusing. Eventually, the solid-like oscillating component of motion disappears; all that remains is the ballistic-collisional motion. That disappearance, a qualitative change in particle dynamics, corresponds to the smooth crossover at the Frenkel line showing the disappearance of the medium-range order correlations at high temperatures \cite{boljcp}. A qualitative change in particle structure corresponds to disappearance of the second  peak in $g(r)$ profile ($g_{CC}(r)$, $g_{CO}(r)$), see Figs. 2c--2d, the same was seen in one-component Lennard-Jones (LJ) supercritical fluid \cite{boljcp}. Unconventionally, $g_{OO}(r)$ (see Figs. 2a--2b) exhibits unusual behaviour: medium-range order correlations remain robust and prevail over the short-range order within very wide temperature range. This effect we attribute to localisation of transverse-like phonon modes which we will discuss in detail below.
In order to explore the microscopic structure of supercritical CO$_2$, we calculate partial static structure factors $S(q)$ from corresponding PDFs. The total structure factor can be decomposed as a sum of partial contributions. These contributions are functions of the various properties which characterize the material. Several methods exist to decomposed the total $S(q)$. The way, we used to calculate the partial structure factors in the present paper, first has been proposed by Faber and Ziman \cite{faber}. In this approach the structure factor is decompose following the correlations between the different chemical species. To describe the correlation between the $\alpha$ and the $\beta$ chemical species the partial structure factor S$^{FZ}_{\alpha\beta}(q)$ is defined by:
\begin{eqnarray}
S^{FZ}_{\alpha\beta}(q) = 1 + 4\pi\varrho \int_{{0}}^{{R_{max}}}dr r^2{\frac{{\sin{qr}}}{{qr}}} \left({g_{\alpha\beta}(r)-1}\right)
\end{eqnarray}
where the $g_{\alpha\beta}(r)$ are the partial pair distribution functions and $R_{max}=$ 20 $\AA$ we used in our calculations. 
\begin{figure}[htp]
  \centering
  \begin{tabular}{cc}
    
    \includegraphics[width=43mm]{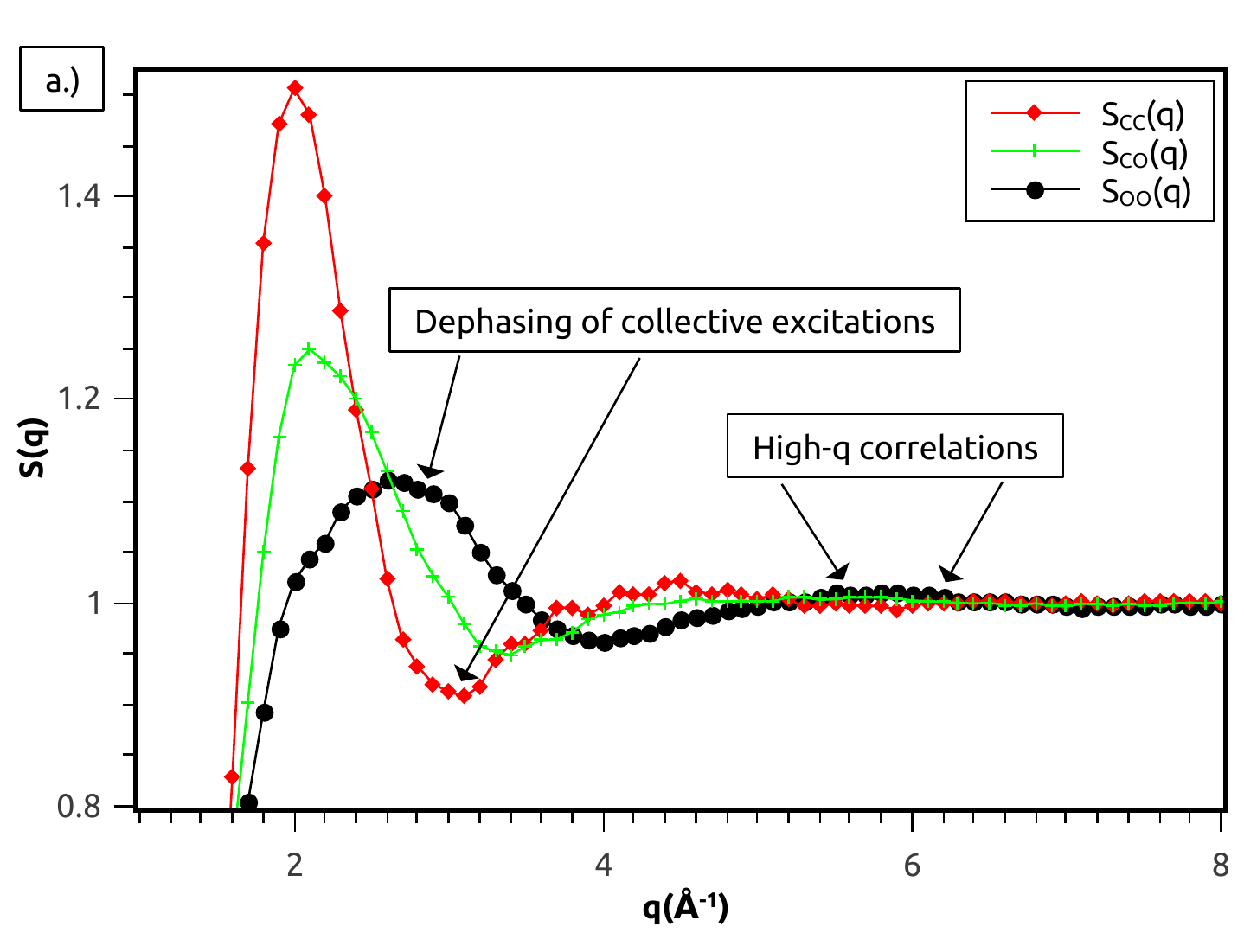}&

    \includegraphics[width=43mm]{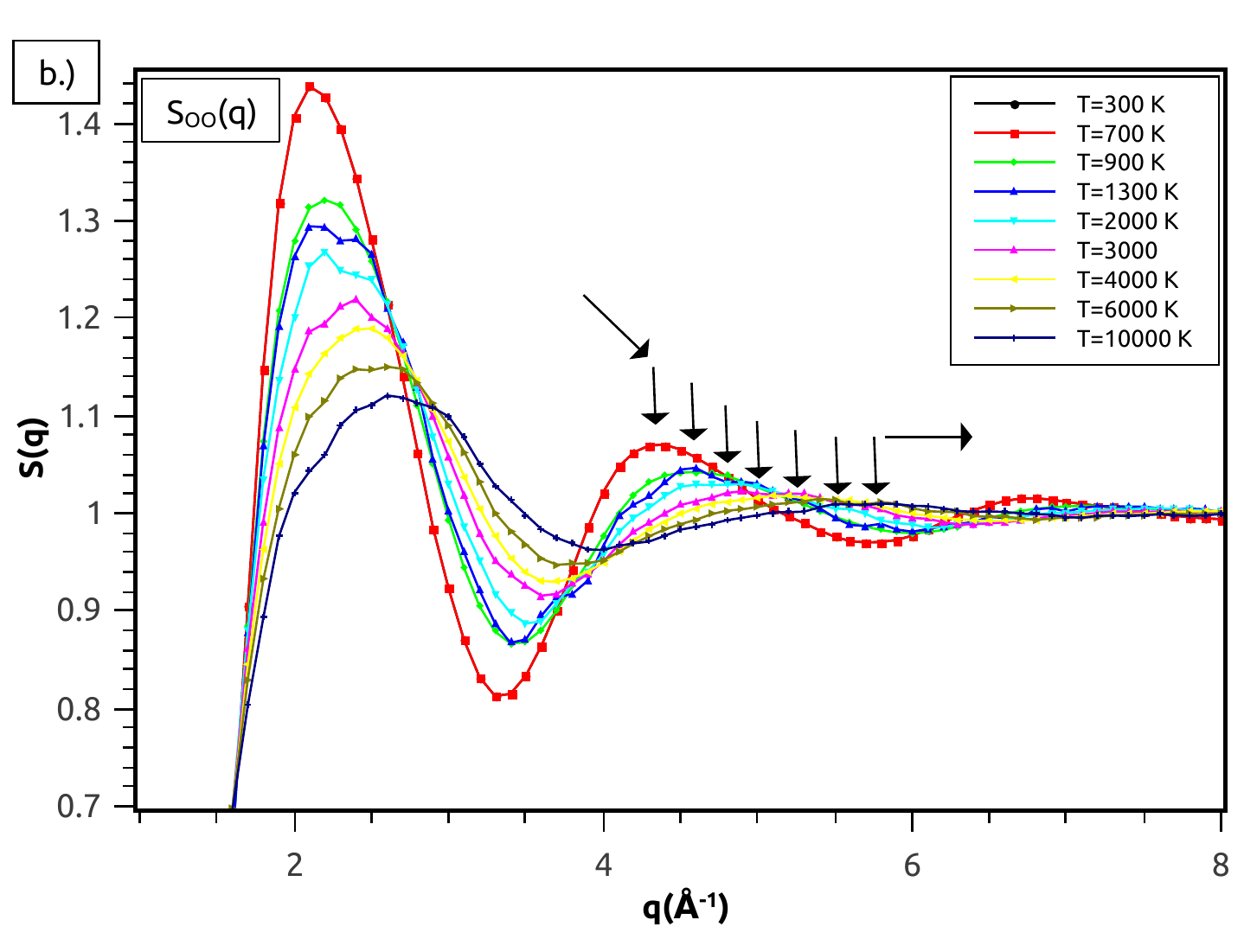}\\

    \includegraphics[width=43mm]{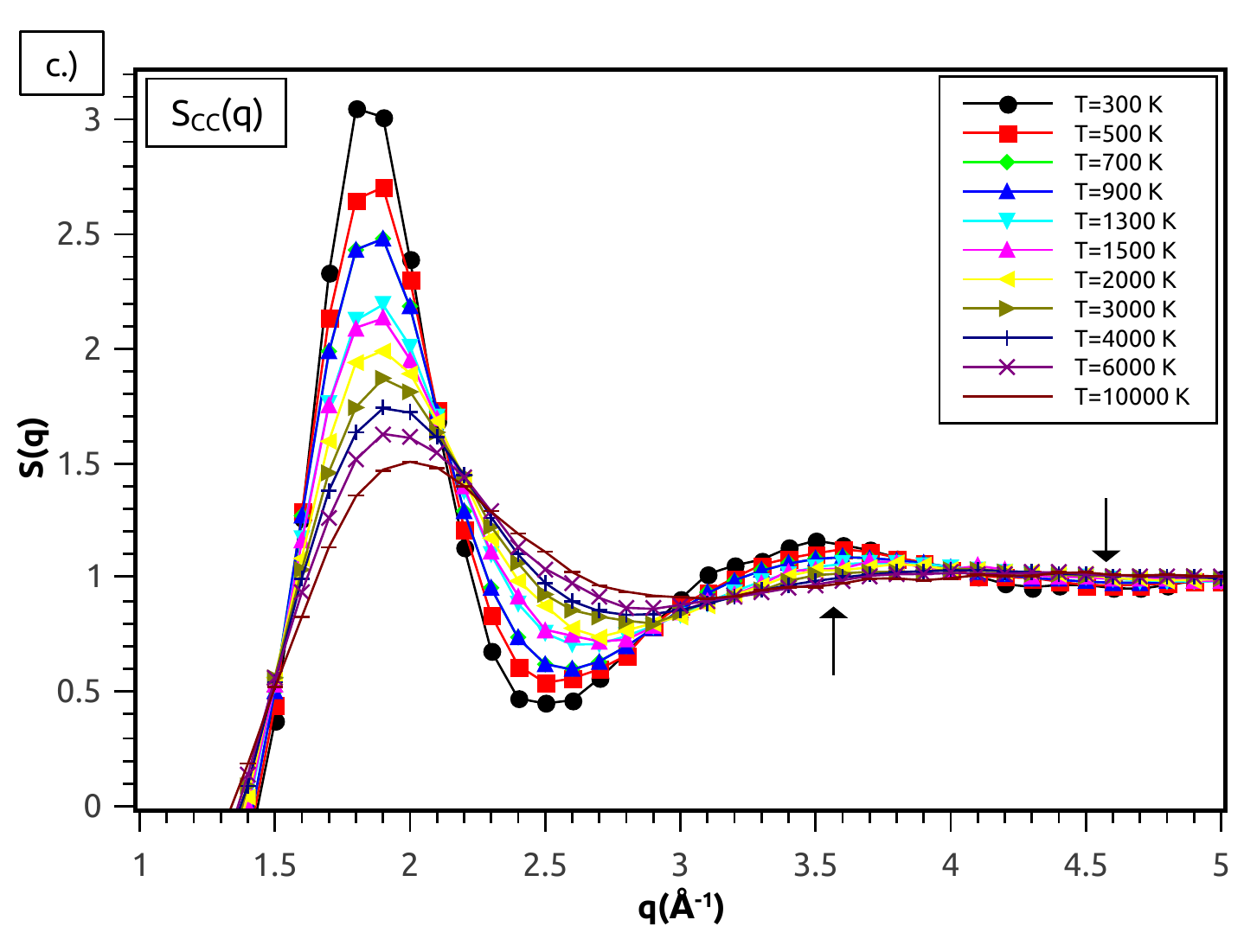}&

    \includegraphics[width=43mm]{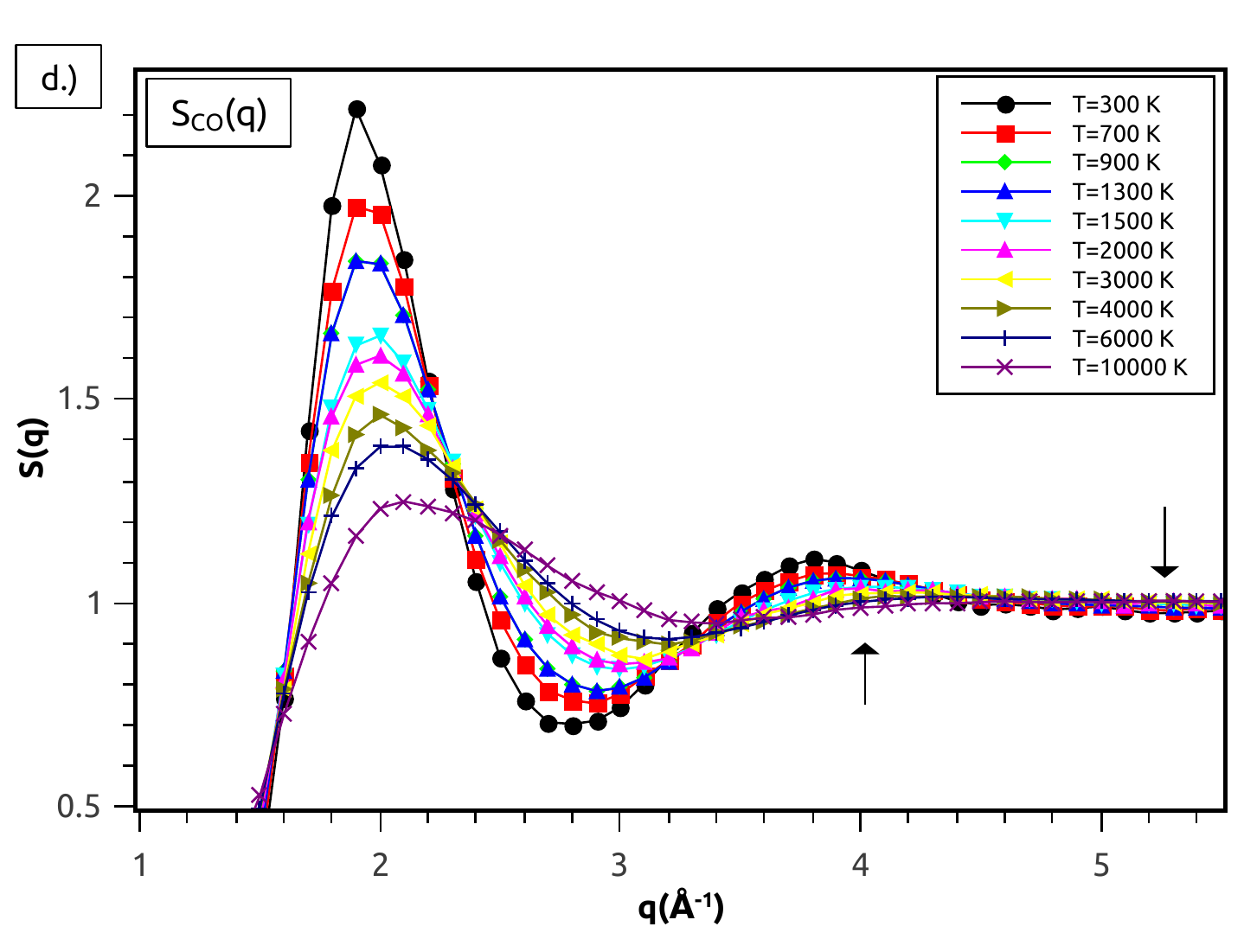}\\

  \end{tabular}
 \caption{ Structural evolution in reciprocal space. Evolution of the partial structure factors.}
 \label{figur}
\end{figure}

In the limiting case of no interaction, the system is an ideal gas and the structure factor is completely featureless: $S(q)=1$. because there is no correlation between the positions $\mathbf{r}_j$ and $\mathbf{r}_k$ of different particles (they are independent random variables), so the off-diagonal terms in equation: 
\begin{eqnarray}
S(q) = 1 + \frac{1}{N} \left \langle \sum_{j \neq k} \mathrm{e}^{-i \mathbf{q} (\mathbf{r}_j - \mathbf{r}_k)} \right \rangle
\end{eqnarray}
average to zero
\begin{eqnarray}
\langle \exp [-i \mathbf{q} (\mathbf{r}_j - \mathbf{r}_k)]\rangle = \langle \exp (-i \mathbf{q} \mathbf{r}_j) \rangle \langle \exp (i \mathbf{q} \mathbf{r}_k) \rangle = 0
\end{eqnarray}
Even for interacting particles, at high scattering vector the structure factor goes to 1. This result follows from equation: $S(q) = 1 + \rho \int_V \mathrm{d} \mathbf{r} \, \mathrm{e}^{-i \mathbf{q}\mathbf{r}} g(r)$,
since $S(q)-1$ is the Fourier transform of the "regular" function $g(r)$ and thus goes to zero for high values of the argument $q$. This reasoning does not hold for a perfect crystal, where the distribution function exhibits infinitely sharp peaks. 
\begin{figure*}[htp]
  \centering
  \begin{tabular}{cc}
    
    \includegraphics[width=48mm]{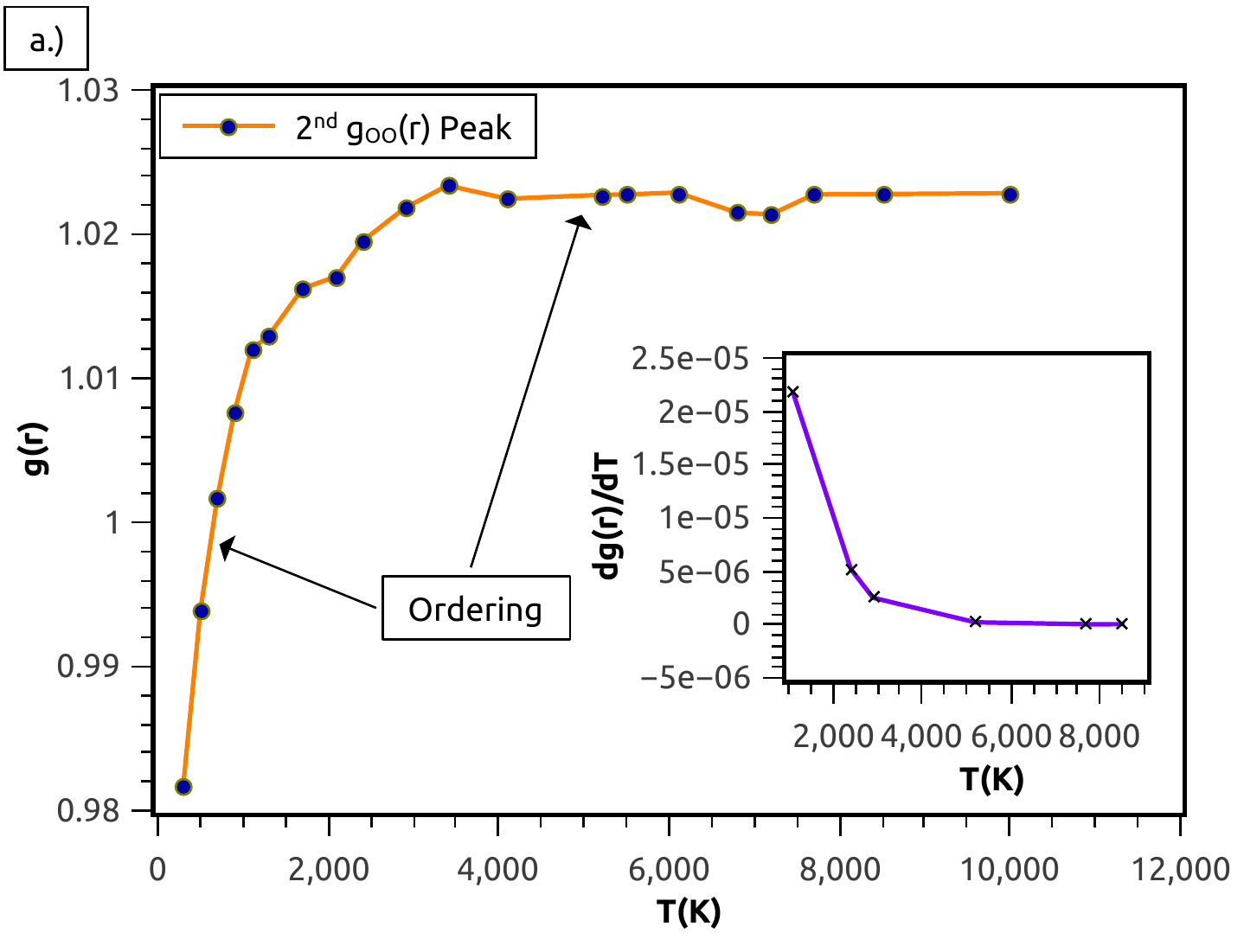}&

    \includegraphics[width=48mm]{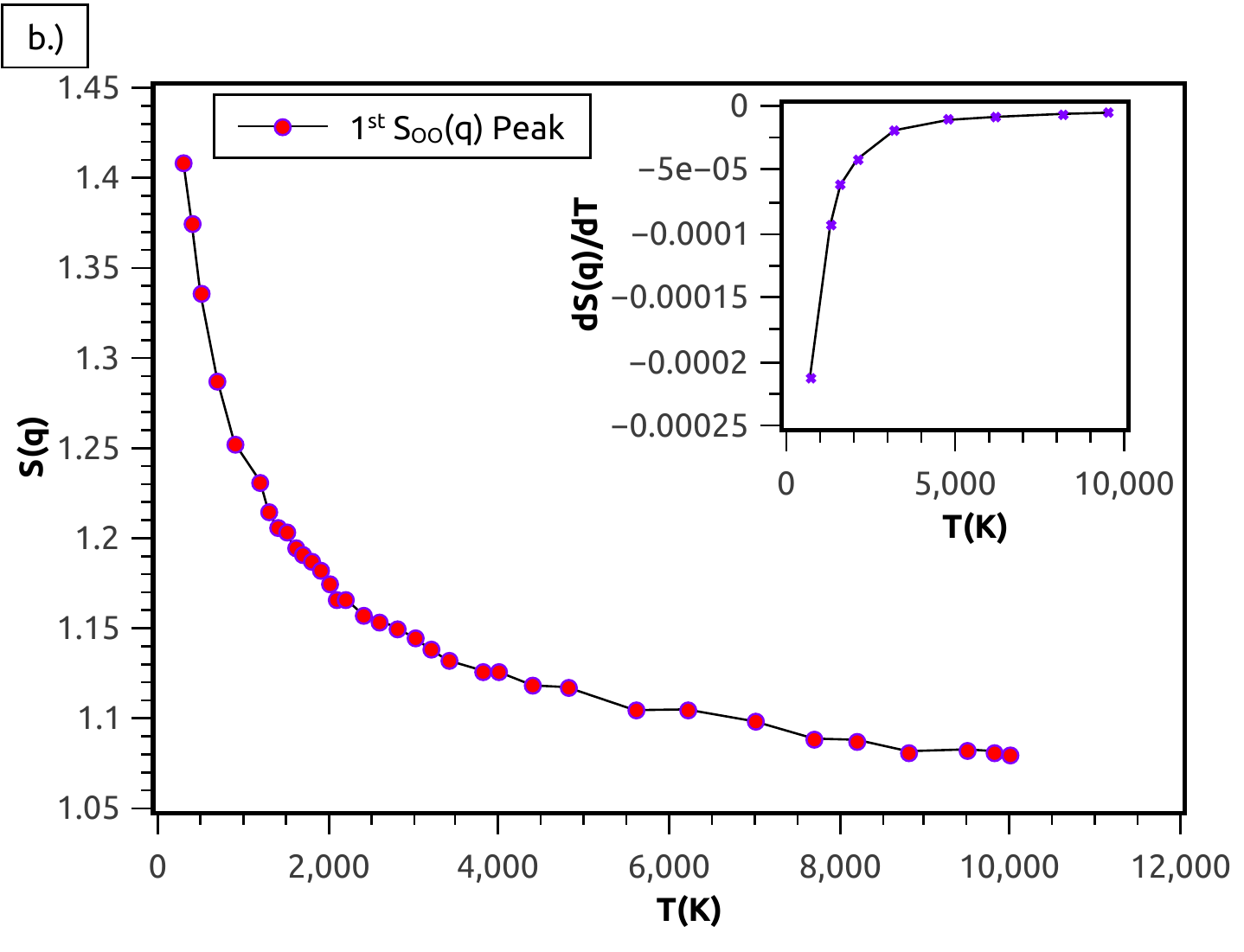}\\

  \end{tabular}
\caption{Ordering within disorder. Evolution of $g_{OO}(r)$ and $S_{OO}(r)$ peaks and their derivatives.}
 \label{figur}
\end{figure*}

The whole shape of the nearest-neighbor intermolecular correlations, in the supercritical CO$_2$, also seem to transform representing a change in the ratio of the peak heights, positions and widths of the O--O, C--C and C--O partial static structure factors, see Figs. 3a--3d. Importantly, $S(q)$ peaks are predicted to change differently with temperature in0 this picture. Indeed, in the low temperature regime, $S_{OO}(q)$, $S_{CC}(q)$ and $S_{CO}(q)$ peaks decrease rapidly and undergo the smooth crossover, clearly seen in Fig. 3b. On the other hand, in the high temperature regime correlations are less sensitive to temperature increase because the dynamics is already randomized by ballistic motions as in a gas  while  $S_{OO}(q)$ also exhibits collective excitations at higher $q$-values retaining the static structure profile as in a liquid. 

As we mentioned previously, the unusual behaviour of $g_{OO}(r)$ discovered above leads to a very important implication: the system exhibits medium-range order correlations. It has been confirmed before \cite{pilgrim1,sette1,inui1}, when liquid relaxation time  $\tau$ (the average time between two consecutive atomic jumps at one point in space \cite{frenkel,bolsr,bolprb,boljap}) approaches its minimal value $\tau_{\rm{D}}$, the Debye vibration period, the system loses the ability to support propagating high-frequency shear modes with $\omega>\frac{2\pi}{\tau}$ and behaves like a gas \cite{bolnatcom}. The inability to support propagating shear modes in a system reflects in the absence of the second $g_{CC}(r)$ and $g_{CO}(r)$ peaks (see Figs. 2c--2d), while the presence of the second $g_{OO}(r)$ peak at high temperatures is attributed to robust localised transverse-like phonon excitations making the supercritical carbon dioxide non-uniform on an intermediate length scale. The persistent local order heterogeneity can be evidenced by the calculation of $g_{OO}(r)$ and $S_{OO}(r)$ peaks (see Figs. 4a--4b), the new effect not hitherto anticipated, in view of the currently perceived physical uniformity and homogeneity of the supercritical state. 

Importantly, we highlight a strong correlation between the atomic structure evolution of
first and second coordination shells with oxygen atoms presented within wide temperature range (see Figs. 2a--2b). The positions of both shells are consistently change with temperature. The number of oxygen atoms in the second shell (see Fig. 4a) persistently remains the same at high temperature while the number of atoms in the first shell constantly reduces due to diffusion. Therefore, atoms inside the nearest neighbor heterogeneity shell play a catalytic role  in provi
The results reported thus far in this work  do not only provide a crucial evidence of additional thermodynamic boundary in the supercritical regime of carbon dioxide. Our observations lead to very intriguing implications for the field of planetary science, specifically for studies of Venus' atmosphere which mainly consist of CO$_2$. 

The planet Venus, now unbearably hot and dry, may have once been far more like Earth, with oceans and continents. Our knowledge concerning the surface of Venus comes from a limited amount of information obtained by the series of landers and probes, and primarily from extensive radar imaging of the planet. Venus is a planet that is similar to Earth in terms of some significant planetary parameters (size, mass, presence of atmosphere) and different in terms of other, equally important ones (absence of an intrinsic magnetic field, large atmospheric mass, carbon dioxide composition of the atmosphere, lack of water, very high surface pressure and temperature). Also the surface of Venus has similar geological features found on Earth: canyons, volcanoes, rift valleys, river-like beds, mountains, craters, and plains. Such features are believed to originate from volcano eruptions and lava flows however there is no direct proof of this hypothesis, which also does not explain the creation of channels of thousand kilometers long in Venusian crust in which lava supposedly selectively flowed. 
\begin{figure}
\includegraphics[scale=0.39]{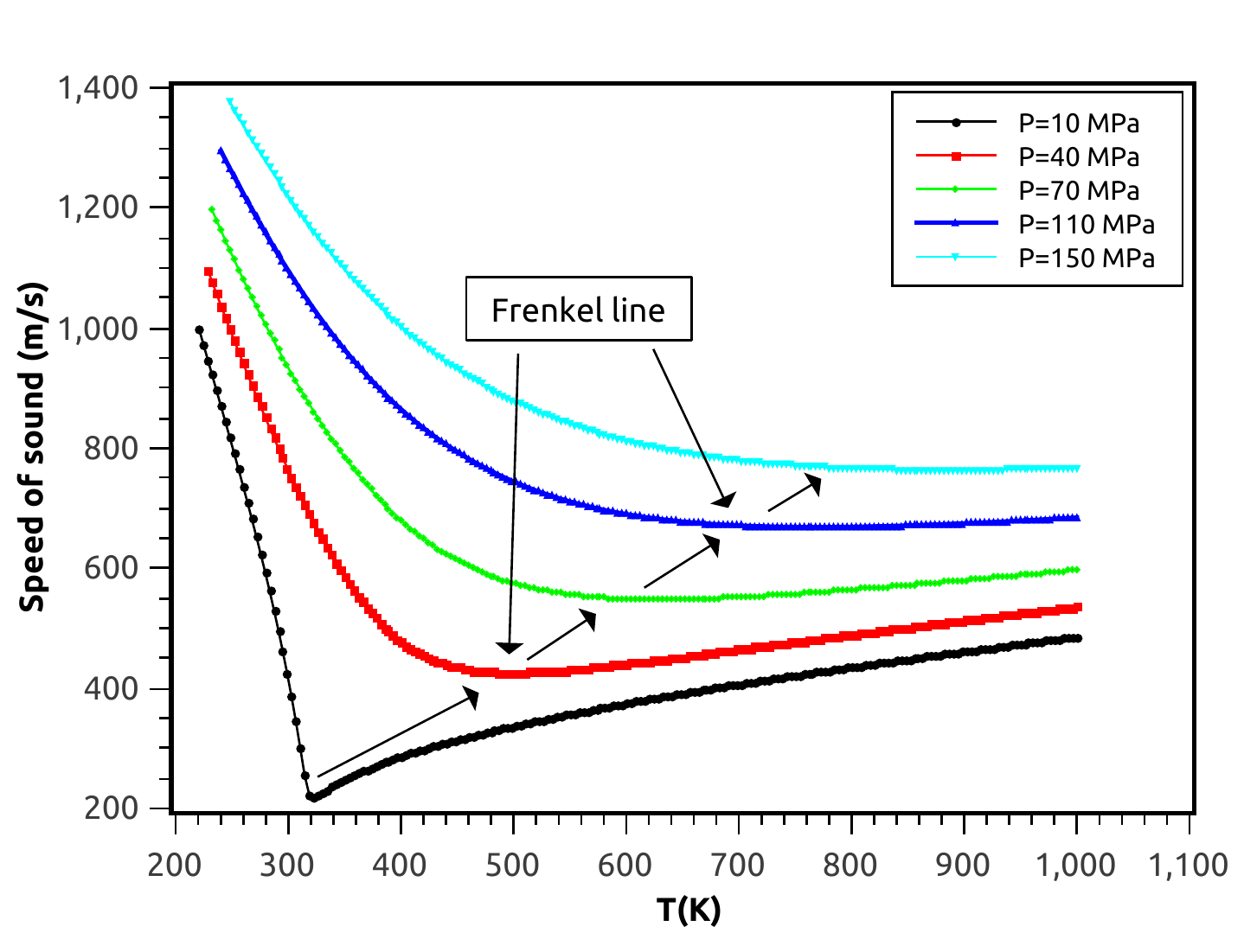}
\caption{{Representative dependencies of the speed of sound as a function of temperature}. Supercritical  carbon dioxide at different pressures. The data are from the NIST database.}
	\label{fig1}
\end{figure}
\begin{figure}
\includegraphics[scale=0.11]{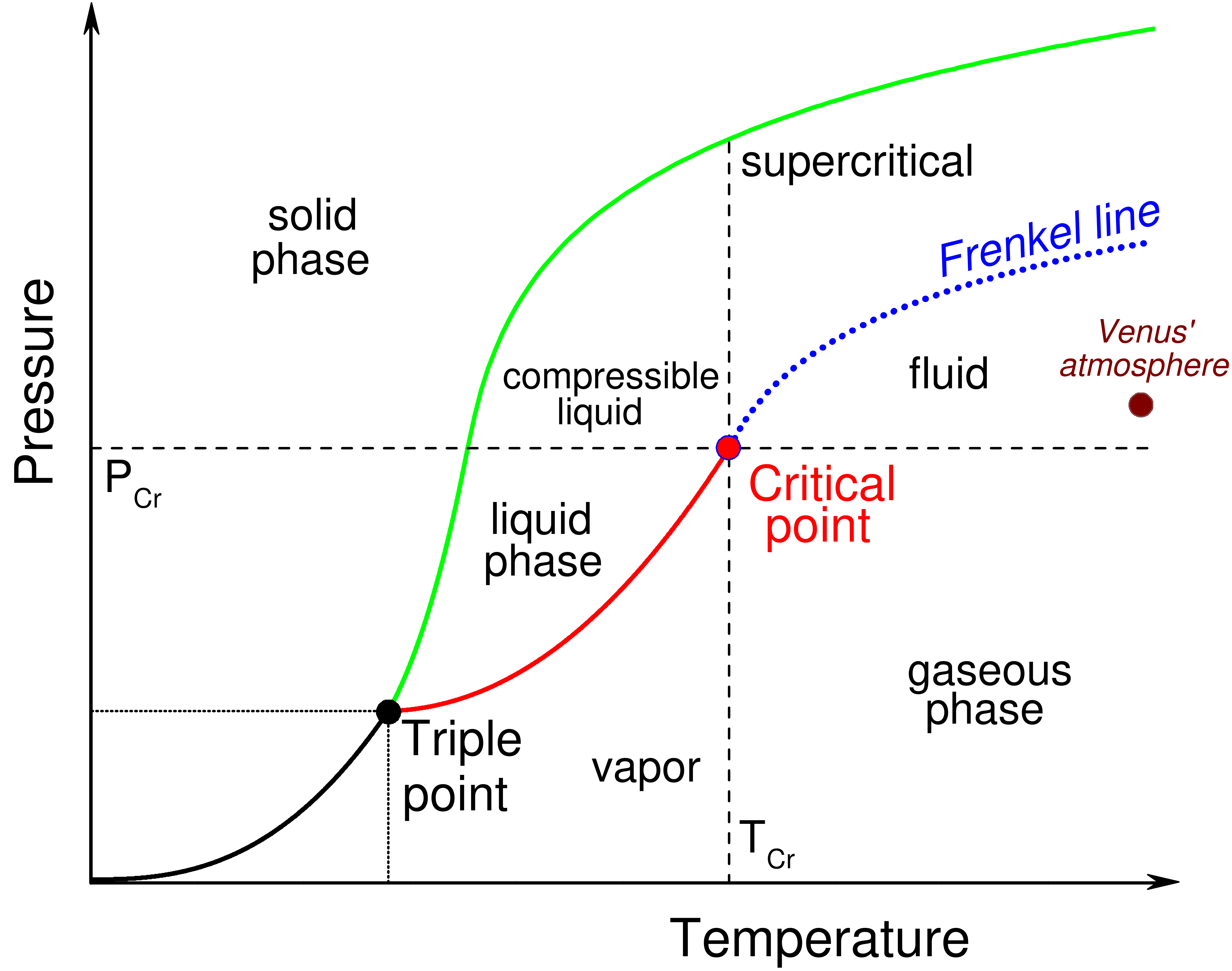}
\caption{{Carbon dioxide pressure-temperature phase diagram}. The surface pressure is 9.3 MPa and the surface temperature is 735 K, above the critical points of both major constituents and making the Venus surface atmosphere a supercritical fluid.}
	\label{fig1}
\end{figure}

Presently, the atmosphere of Venus is mostly carbon dioxide, 96.5\% by volume. It is believed that in the past Venus may have had enough water in the atmosphere to cover the entire planet by an ocean 25 meters deep. However, even if there was water it was probably too warm for it to fall as rain and form oceans and thus much of the water remained in the atmosphere as well as other constituents including carbon dioxide. Model calculations show that the extensive volcanic activity that occurred during the early era might have also increased the abundance of gases like H$_2$O and SO$_2$ in the atmosphere, strengthening the greenhouse effect \cite{hashimoto}. Although currently, the surface pressure on Venus is 9.3 MPa and the surface temperature is 735 K, because of abundance of CO$_2$, H$_2$O and SO$_2$ the early surface pressure may have been a few tens of MPa \cite{bookvenus}. At the same time, these conditions may have been accompanied by significant (± 300 K) and rather long duration (100-200 million years) changes in temperature and pressure making it possible for liquid CO$_2$ to be formed in the vicinity of the Venus surface at very early stages of greenhouse effect era. To determine the location of the Frenkel line in CO$_2$ (see Figs. 5a-5b) at pressure and temperature conditions we apply the same analysis as we used in our previous work for H$_2$ \cite{trachenko}. These conditions correspond to CO$_2$ above the Frenkel line where it exhibits liquid-like behavior. This in turn makes it plausible that such geological features like rift valleys, river-like beds and plains are the fingerprints of near-surface activity of liquid-like supercritical CO$_2$.

Although the structural order of a fluid is usually enhanced by isothermal compression or isochoric cooling, a few notable systems show the opposite behaviours. Specifically, increasing density can disrupt the structure of liquid-like fluids, while lowering temperature or strengthening of attractive interactions can weaken the correlations of fluids with short-range attractions. It is a particularly insightful quantity to study because its contributions from the various coordination shells of the pair distribution function can be readily determined, and it correlates strongly with self-diffusivity, which allows it to provide insights into the dynamic and structural anomalies of fluids.

In this work, we report on the carbon dioxide heterogeneity shell structure where, in the first shell, both carbon and oxygen atoms experience gas-like type interactions with short range order correlations, while within the second shell oxygen atoms essentially exhibit liquid-like type of interactions with medium range order correlations due to localisation of transverse-like phonon packets. Therefore, the local order heterogeneity remains in the three phase-like equilibrium within very wide temperature range. Atoms inside the nearest neighbor heterogeneity shell play a catalytic role due to short range order interactions, providing a mechanism for diffusion in the supercritical carbon dioxide on an intermediate length scale. Hence, persistent local order heterogeneities in the supercritical carbon dioxide, due to their peculiar structures, might selectively swell particular nanodomains of block copolymers to produce nanocellular and nanoporous structures more efficiently than it was seen before.

The emergence of local order heterogeneities and existence of localised transverse-like phonon modes are closely inter-related. From the phonon theory of liquid thermodynamics \cite{bolsr} we know that liquids support high-frequency propagating shear phonon modes and lose this ability with temperature increase well extending into the supercritical region. Oxygen atoms in the second shell define a boundary between propagating and localized phonon modes in the supercritical carbon dioxide. 6--8 $\AA$ value is estimated length for the highly localized  transverse-like phonon modes and also corresponds to the size of the heterogeneities where the phonon packets are localized (see Figs. 1--2). Similarly, propagating length of the transverse acoustic modes were determined to be 4--10 $\AA$ corresponding to the size of cages formed instantaneously in liquid Ga \cite{inui1} and liquid Sn \cite{hosokawa1}. Thus, the supercritical state is also universally amenable to supporting fundamental interlinks between fundamental system properties such as structure and dynamics.

In summary, the results of this study support the emerging view that the supercritical state is non-uniform and heterogeneous which can generally be attributed to quantifiable structural changes in the first and higher coordination shells of the partial pair distribution functions and their corresponding static structure factors.  Away from the critical point, the structure of the PDFs typically preserves for distances comparable to a few particle diameters, reflecting the short range of the inter-particle correlations. The existence of persistent medium-range order correlations in the supercritical CO$_2$ has not been hitherto anticipated, and is contrary to how the supercritical state has been viewed until now. 

Our results call for experimental observation of the persistent local order heterogeneities. This observation would not only prove the existence of additional thermodynamic boundary in the supercritical regime, it would also provide a compelling evidence that supercritical liquid-like carbon dioxide may have influenced the formation  of current geological map of Venus.


{\it Acknowledgements.} Dima Bolmatov thanks Ben Widom and Cornell University for support. We are grateful to Neil Ashcroft and Martin Dove for discussions.

\end{document}